\documentclass[12pt]{article}
\begin{document}

\author{C. Bizdadea\thanks{%
e-mail address: bizdadea@central.ucv.ro},
E. M. Cioroianu\thanks{%
e-mail address: manache@central.ucv.ro},
M. T. Miaut\u {a}, \\
I. Negru\thanks{%
e-mail address: inegru@central.ucv.ro},
S. O. Saliu\thanks{%
e-mail address: osaliu@central.ucv.ro} \\
Faculty of Physics, University of Craiova\\
13 A. I. Cuza Str., Craiova RO-1100, Romania}
\title{Lagrangian cohomological couplings among
vector fields and matter fields}
\maketitle

\begin{abstract}
Consistent couplings between a set of vector
fields and a system of matter
fields are analysed in the framework of
Lagrangian BRST cohomology.

\end{abstract}

\section{Introduction}

The antifield-BRST symmetry \cite{1}--\cite{2}
viewed from a cohomological
perspective provided a helpful instrument
for analysing consistent
interactions in gauge theories \cite{3}--\cite{6}.
In this light, many
models of interest in theoretical physics, like,
for example, the Yang-Mills
theory \cite{7}, the Chapline-Manton model \cite{8},
$p$-forms and chiral $p$%
-forms \cite{9}--\cite{13}, as well as nonlinear
gauge theories \cite{13a},
have been derived along the deformation of the
master equation. Also, it is
important to notice the deformation results
connected to Einstein's gravity
theory \cite{14} and four- and eleven-dimensional
supergravity \cite{15}.
Meanwhile, the problem of obtaining consistent
deformations has naturally
found its extension at the Hamiltonian level
\cite{16}--\cite{19}.

In this paper we investigate consistent
Lagrangian couplings that can be
added between a set of vector fields and a
system of matter fields (of spin
0, respectively, 1/2) by means of the
deformation of the master equation.
This approach represents an extension of
our former results exposed in \cite
{20a} on the abelian case. Our treatment
goes as follows. We begin with a
``free'' action written as the sum between
the action for a set of vector
fields and an action describing a matter theory,
and construct the
corresponding ``free'' Lagrangian BRST
differential $s$, which is found to
decompose as the sum between the Koszul-Tate
differential and the exterior
derivative along the gauge orbits,
$s=\delta +\gamma $. Further, we analyse
the deformation of the associated solution
to the master equation. The
first-order deformation belongs to
$H^{0}\left( s|d\right) $, where $%
d=dx^{\mu }\partial _{\mu }$ denotes
the exterior space-time derivative. The
computation of the cohomological space
$H^{0}\left( s|d\right) $ proceeds by
expanding the co-cycles according to
the antighost number, and by
subsequently using the cohomological spaces
$H\left( \gamma \right) $ and $%
H\left( \delta |d\right) $. The first-order
deformation reveals two
different types of couplings. One involves
only the vector fields (the cubic
vertex of Yang-Mills theory), and requires
no supplementary assumptions. The
other demands that the cohomological space
$H_{1}\left( \delta |d\right) $
is not empty, or, in other words, that the
matter theory should display some
global invariances that result in some
conserved currents $j_{\;\;a}^{\mu }$%
. Consequently, it follows that the second
type of couplings from the
first-order deformation is written in the
form $j_{\;\;a}^{\mu }A_{\mu }^{a}$%
. The consistency of the first-order
deformation asks that the deformed
gauge algebra is Lie, and, moreover,
outputs precisely the well-known
quartic vertex of Yang-Mills theory.
Concerning the higher-order
deformations, there appear two main situations.
If the conserved currents $%
j_{\;\;a}^{\mu }$ transform under the
gauge version of the rigid symmetries
of the matter fields according to the
adjoint representation of the Lie
gauge algebra, then all the other
deformations involving the matter fields,
of order two and higher, vanish. In the
opposite case, at least the
second-order deformation implying matter
fields is non-vanishing, but in
principle there might be other non-trivial terms.

The paper is structured in five sections.
In Sec. 2 we briefly review the
problem of deformation of the master
equation. Sec. 3 focuses on the
analysis of the consistent couplings
between a set of vector fields and a
collection of matter fields by means
of deforming the solution to the master
equation. In Sec. 4 we apply our treatment
to two cases of interest, where
the role of the matter fields is played
by a set of real scalar fields,
respectively, by a collection of Dirac
fields. Sec. 5 ends the paper with
some conclusions.

\section{Deformation of the master
equation: a brief review}

We begin with a ``free'' gauge theory,
described by an action $S_{0}\left[
\Phi ^{\alpha _{0}}\right] $, invariant
under some gauge transformations 
\begin{equation}
\delta _{\epsilon }\Phi ^{\alpha _{0}}=
Z_{\;\;\alpha _{1}}^{\alpha
_{0}}\epsilon ^{\alpha _{1}},\;
\frac{\delta S_{0}}{\delta \Phi ^{\alpha _{0}}%
}Z_{\;\;\alpha _{1}}^{\alpha _{0}}=0,
\label{2.1}
\end{equation}
and consider the problem of consistent
interactions that can be introduced
among the fields $\Phi ^{\alpha _{0}}$,
so the couplings preserve the
original number of gauge symmetries.
This matter is addressed by means of
reformulating the problem of constructing
consistent interactions as a
deformation problem of the solution to the
master equation corresponding to
the ``free'' theory \cite{3}. Such a
reformulation is possible due to the
fact that the solution to the master
equation contains all the information
on the gauge structure of the theory.
If a consistent interacting gauge
theory can be constructed, then the
solution $\tilde{S}$ to the master
equation associated with the ``free''
theory can be deformed into a solution
$S$,
\begin{eqnarray}
&&\tilde{S}\rightarrow S=\tilde{S}+gS_{1}+
g^{2}S_{2}+\cdots =  \nonumber \\
&&\tilde{S}+g\int d^{D}x\,a+g^{2}\int d^{D}x\,b+
\cdots ,  \label{2.2}
\end{eqnarray}
of the master equation for the deformed theory
\begin{equation}
\left( S,S\right) =0,  \label{2.3}
\end{equation}
that displays the same ghost and antifield
spectra. According to the
deformation parameter $g$, eq. (\ref{2.3}) splits into
\begin{equation}
\left( \tilde{S},\tilde{S}\right) =0,  \label{2.4}
\end{equation}
\begin{equation}
2\left( S_{1},\tilde{S}\right) =0,  \label{2.5}
\end{equation}
\begin{equation}
2\left( S_{2},\tilde{S}\right) +
\left( S_{1},S_{1}\right) =0,  \label{2.6}
\end{equation}
\begin{equation}
\left( S_{3},\tilde{S}\right) +
\left( S_{1},S_{2}\right) =0,  \label{2.7}
\end{equation}
\[
\vdots
\]
While eq. (\ref{2.4}) is fulfilled by
hypothesis, the next one requires that
$S_{1}$ is a co-cycle of the ``free''
BRST differential $s\bullet =\left(
\bullet ,\tilde{S}\right) $. However,
only cohomologically non-trivial
solutions to (\ref{2.5}) should be
taken into account, as the BRST-exact
ones can be eliminated by a (in general non-linear)
field redefinition, such
that $S_{1}$ pertains to the ghost number
zero cohomological space of $s$, $%
H^{0}\left( s\right) $, which is known to be
isomorphic to the space of
physical observables of the ``free'' theory.
It has been shown in \cite{3},
\cite{20} on behalf of the triviality of the
antibracket in the cohomology
that there are no obstructions in finding
solutions to the remaining eqs. ((%
\ref{2.6}--\ref{2.7}), etc.). Unfortunately,
the resulting interactions may
not be local, and there might appear
obstructions if one insists on the
locality of the deformations. The analysis
of these obstructions can be done
with the help of cohomological techniques.
However, as it will be seen below
(see Sec. 4), the interaction terms to be
dealt with turn out to be local.

\section{Couplings among vector fields and
matter fields from BRST cohomology}

The scope of this section is to analyse the
consistent interactions among
vector fields and matter fields with the help
of the Lagrangian BRST
deformation procedure.

\subsection{Free BRST differential}

We start from a ``free'' theory whose
Lagrangian action is written as the
sum between the action for a collection
of vector fields $A_{\mu }^{a}$ and
an action involving some matter fields $y^{i}$
\begin{equation}
\tilde{S}_{0}^{L}\left[ A_{\mu }^{a},y^{i}\right] =
\int d^{D}x\left( -\frac{1%
}{4}F_{\mu \nu }^{a}F_{a}^{\mu \nu }+
{\mathcal{L}}_{0}\left( y^{i},\partial
_{\mu }y^{i},\ldots ,
\partial _{\mu _{1}}\cdots \partial _{\mu
_{k}}y^{i}\right) \right) ,  \label{4.1}
\end{equation}
where the field strength is defined as
$F_{\mu \nu }^{a}=\partial _{\mu
}A_{\nu }^{a}-\partial _{\nu }A_{\mu }^{a}$.
The Grassmann parity of a
matter field $y^{i}$ is denoted by
$\varepsilon _{i}$. We assume that the
matter fields possess no gauge
invariances of their own, hence action (\ref
{4.1}) is invariant under the gauge
transformations
\begin{equation}
\delta _{\epsilon }A_{\mu }^{a}=
\partial _{\mu }\epsilon ^{a},\;\delta
_{\epsilon }y^{i}=0.  \label{4.2}
\end{equation}
Then, the solution to the master equation
for this ``free'' theory reads as
\begin{equation}
\tilde{S}=\tilde{S}_{0}^{L}\left[
A_{\mu }^{a},y^{i}\right] +\int
d^{D}x\,A_{a}^{*\mu }\partial _{\mu }
\eta ^{a},  \label{4.3}
\end{equation}
where $\eta ^{a}$ are the fermionic ghosts,
and the star variables represent
the antifields of the corresponding
fields/ghosts. The antifields $%
A_{a}^{*\mu }$ of the vector fields are
fermionic, while those of the
ghosts, $\eta _{a}^{*}$, are bosonic.
The grading of the BRST differential
is named ghost number ($\mathrm{gh}$),
and is defined like the difference
between the degree of the exterior
derivative along the gauge orbits, pure
ghost number ($\mathrm{pgh}$), and
the grading of the Koszul-Tate
differential, antighost number
($\mathrm{antigh}$), where
\begin{equation}
\mathrm{pgh}\left( A_{\mu }^{a}\right) =
\mathrm{pgh}\left( y^{i}\right) =%
\mathrm{pgh}\left( A_{a}^{*\mu }\right) =
\mathrm{pgh}\left( y_{i}^{*}\right)
=\mathrm{pgh}\left( \eta _{a}^{*}\right) =
0,\;\mathrm{pgh}\left( \eta
^{a}\right) =1,  \label{4.3a}
\end{equation}
\begin{equation}
\mathrm{antigh}\left( A_{\mu }^{a}\right) =
\mathrm{antigh}\left(
y^{i}\right) =0,\;\mathrm{antigh}
\left( A_{a}^{*\mu }\right) =\mathrm{antigh}%
\left( y_{i}^{*}\right) =1,  \label{4.3b}
\end{equation}
\begin{equation}
\mathrm{antigh}\left( \eta _{a}^{*}\right) =
2,\;\mathrm{antigh}\left( \eta
^{a}\right) =0.  \label{4.3c}
\end{equation}
The BRST symmetry of the ``free'' theory, $s\bullet =
\left( \bullet ,\tilde{S%
}\right) $, simply decomposes as the
sum between the Koszul-Tate
differential and the exterior longitudinal derivative
\begin{equation}
s=\delta +\gamma ,  \label{4.4}
\end{equation}
where these operators act on the generators of
the BRST complex through the
relations 
\begin{equation}
\delta A_{\mu }^{a}=0,\;\delta y^{i}=
0,\;\delta \eta ^{a}=0,  \label{4.5}
\end{equation}
\begin{equation}
\delta A_{a}^{*\mu }=
-\partial _{\nu }F_{a}^{\nu \mu },\;\delta y_{i}^{*}=-%
\frac{\delta ^{L}{\mathcal{L}}_{0}}{\delta y^{i}},
\;\delta \eta _{a}^{*}=
-\partial _{\mu }A_{a}^{*\mu },  \label{4.6}
\end{equation}
\begin{equation}
\gamma A_{\mu }^{a}=
\partial _{\mu }\eta ^{a},\;\gamma y^{i}=0,\;\gamma \eta
^{a}=0,  \label{4.7}
\end{equation}
\begin{equation}
\gamma A_{a}^{*\mu }=0,\;\gamma y_{i}^{*}=
0,\;\gamma \eta _{a}^{*}=0,
\label{4.8}
\end{equation}
which will be used in the sequel during
the deformation process.

\subsection{First-order deformations}

For determining the consistent interactions
that can be added to the
``free'' gauge theory under discussion,
we examine the eqs. (\ref{2.5}--\ref%
{2.7}), etc., by relying on the action
of the ``free'' BRST differential.
Using the notations from (\ref{2.2}),
the local form of the eq. (\ref{2.5})
is
\begin{equation}
sa=\partial _{\mu }n^{\mu },  \label{4.9}
\end{equation}
for some local $n^{\mu }$, and it holds
if and only if $a$ is a $s$-co-cycle
modulo $d$. In order to solve this eq.,
we develop $a$ according to the
antighost number,
\begin{equation}
a=a_{0}+a_{1}+\cdots +a_{j},  \label{4.9a}
\end{equation}
where $\mathrm{antigh}\left( a_{i}\right) =i$,
and the last term in $a$ can
be assumed to be annihilated by
$\gamma $, $\gamma a_{j}=0$. Thus, we need
the cohomology of $\gamma $,
$H\left( \gamma \right) $, in order to
determine the components of highest
antighost number in $a$. From (\ref{4.7}-%
\ref{4.8}) it is simple to see that
the cohomology of $\gamma $ is spanned
by $F_{\mu \nu }^{a}=
\partial _{\left[ \mu \right. }A_{\left. \nu \right]
}^{a}$, $y^{i}$ and their derivatives,
by the antifields and their
derivatives, as well as by the
undifferentiated ghosts $\eta ^{a}$. (The
derivatives of the ghosts do not bring
any contribution to $H\left( \gamma
\right) $ as they are $\gamma $-exact.)
If we denote by $e^{M}\left( \eta
^{a}\right) $ a basis in the space of
the polynomials in the ghosts, it
follows that the general solution to the
equation $\gamma \alpha =0$ (up to
a trivial term) takes the form
\begin{equation}
\alpha =\alpha _{M}\left( \left[
F_{\mu \nu }^{a}\right] ,\left[
y^{i}\right] ,\left[ A_{a}^{*\mu }\right] ,
\left[ y_{i}^{*}\right] ,\left[
\eta _{a}^{*}\right] \right)
e^{M}\left( \eta ^{a}\right) ,  \label{4.9c}
\end{equation}
where the notation $f\left[ q\right] $
signifies that $f$ depends on $q$ and
its derivatives up to a finite order.
Relying on the fact that $\mathrm{gh}%
\left( a\right) =0$, it results that
$\mathrm{pgh}\left( a_{i}\right) =i$,
hence from (\ref{4.9c}) we have that
the last representative in $a$ is of
the type
\begin{equation}
a_{j}=\frac{1}{\left( j+1\right) !}
a_{a_{1}\cdots a_{j}}\left( \left[ F_{\mu
\nu }^{a}\right] ,\left[ y^{i}\right] ,
\left[ A_{a}^{*\mu }\right] ,\left[
y_{i}^{*}\right] ,\left[ \eta _{a}^{*}
\right] \right) \eta ^{a_{1}}\cdots
\eta ^{a_{j}}.  \label{4.10}
\end{equation}
The eq. (\ref{4.9}) projected on antighost
number $\left( j-1\right) $
becomes
\begin{equation}
\delta a_{j}+\gamma a_{j-1}=
\partial _{\mu }m^{\mu }.  \label{4.10a}
\end{equation}
The last equation possesses solutions if all
$a_{a_{1}\cdots a_{j}}$ pertain
to $H_{j}\left( \delta |d\right) $.
In the meantime, applying the results
from \cite{20}, it follows that
$H_{j}\left( \delta |d\right) $ vanishes for
$j>2$, so we can assume that the
first-order deformation stops after the
first three elements
\begin{equation}
a=a_{0}+a_{1}+a_{2},  \label{4.11}
\end{equation}
with $a_{2}=\frac{1}{2}a_{ab}\eta ^{a}\eta ^{b}$ and
$a_{ab}=-a_{ba}$ from $%
H_{2}\left( \delta |d\right) $, i.e.
\begin{equation}
\delta a_{ab}=\partial _{\mu }k_{ab}^{\mu }.  \label{4.12}
\end{equation}
Nevertheless, the most general element from
$H_{2}\left( \delta |d\right) $
can be represented as
\begin{equation}
\lambda =\lambda ^{a}\eta _{a}^{*},  \label{4.13}
\end{equation}
with constant $\lambda ^{a}$, such that
$\delta \lambda =\partial _{\mu
}\left( -\lambda ^{a}A_{a}^{*\mu }\right) $.
These considerations lead to
\begin{equation}
a_{ab}=-f_{\;\;ab}^{c}\eta _{c}^{*},  \label{4.13a}
\end{equation}
with $f_{\;\;ab}^{c}=-f_{\;\;ba}^{c}$ some constants.
Combining the above
results, we can state that the antighost
number two component of $a$ can be
written in the form
\begin{equation}
a_{2}=-\frac{1}{2}f_{\;\;ab}^{c}
\eta _{c}^{*}\eta ^{a}\eta ^{b}.
\label{4.13b}
\end{equation}
The eq. (\ref{4.9}) projected on antighost
number one becomes $\delta
a_{2}+\gamma a_{1}=\partial _{\mu }k^{\mu }$.
Using (\ref{4.13b}), by direct
computation we find
\begin{equation}
\delta a_{2}=\partial _{\mu }\left(
\frac{1}{2}f_{\;\;ab}^{c}A_{c}^{*\mu
}\eta ^{a}\eta ^{b}\right) -\gamma
\left( f_{\;\;ab}^{c}A_{c}^{*\mu }A_{\mu
}^{b}\eta ^{a}+y_{i}^{*}T_{\;\;a}^{i}
\eta ^{a}\right) ,  \label{4.13c}
\end{equation}
where $T_{\;\;a}^{i}$ are some local
functions of $y^{i}$ and their
derivatives. From (\ref{4.13c}) we read
the antighost number one term of the
first-order deformation
\begin{equation}
a_{1}=f_{\;\;ab}^{c}A_{c}^{*\mu }A_{\mu }^{b}\eta
^{a}+y_{i}^{*}T_{\;\;a}^{i}\eta ^{a}.  \label{4.13d}
\end{equation}
In order to solve the eq. controlling
the antighost number zero deformation,
$\delta a_{1}+\gamma a_{0}=
\partial _{\mu }l^{\mu }$, from (\ref{4.13d}) we
deduce that
\begin{equation}
\delta a_{1}=\partial _{\mu }\left(
f_{\;\;ab}^{c}F_{c}^{\mu \nu }A_{\nu
}^{b}\eta ^{a}\right) -\gamma \left(
\frac{1}{2}f_{\;\;ab}^{c}F_{c}^{\mu \nu
}A_{\mu }^{a}A_{\nu }^{b}\right) +
\left( -\right) ^{\varepsilon _{i}}\frac{%
\delta ^{L}{\mathcal{L}}_{0}}{\delta y^{i}}
T_{\;\;a}^{i}\eta ^{a}.
\label{4.14}
\end{equation}
Thus, for developing a consistent
deformation procedure, it is necessary
that the third term in the right
hand-side of the last equation is $\gamma $%
-exact modulo $d$. This takes place if
\begin{equation}
\left( -\right) ^{\varepsilon _{i}}
\frac{\delta ^{L}{\mathcal{L}}_{0}}{%
\delta y^{i}}T_{\;\;a}^{i}=
\partial _{\mu }j_{\;\;a}^{\mu },  \label{4.14a}
\end{equation}
for some local $j_{\;\;a}^{\mu }$.
The last relation is nothing but
Noether's theorem expressing the
appearance of the on-shell conserved
currents $j_{\;\;a}^{\mu }$ (on-shell
means, as usually, on the stationary
surface of field equations) deriving
from the invariance of the Lagrangian
action of the matter fields under the
global symmetries
\begin{equation}
\Delta y^{i}=T_{\;\;a}^{i}\left(
y^{i},\partial _{\mu }y^{i},\ldots
,\partial _{\mu _{1}}\cdots
\partial _{\mu _{k}}y^{i}\right) \xi ^{a},
\label{4.14b}
\end{equation}
with $\xi ^{a}$ some bosonic constant
parameters. In conclusion, the
consistency of our deformation procedure
requires in the first place that
the matter theory is invariant under
the rigid transformations (\ref{4.14b}%
), which lead to Noether's theorem
(\ref{4.14a}). In the sequel we will
assume that the matter theory indeed
meets this requirement. The eq. (\ref%
{4.14a}) may be rewritten in terms of
the Koszul-Tate differential as
\begin{equation}
\partial _{\mu }j_{\;\;a}^{\mu }=
\delta \left(
-y_{i}^{*}T_{\;\;a}^{i}\right) \equiv
\delta \alpha _{a},  \label{4.14c}
\end{equation}
and it expresses the relationship between
the rigid symmetries (\ref{4.14b})
and the homological space
$H_{1}\left( \delta |d\right) $. Explicitly, it
shows that a global symmetry
(materialized in a conserved current) defines
an element $\alpha _{a}$ of
$H_{1}\left( \delta |d\right) $, i.e., an
element of antighost number equal to one
that is $\delta $-closed modulo $d$%
. A global symmetry is said to be trivial
if the corresponding $\alpha _{a}$
is in a trivial class of
$H_{1}\left( \delta |d\right) $, hence if it is $%
\delta $-exact modulo $d$%
\begin{equation}
\alpha _{a}=\delta b_{a}+
\partial _{\mu }c_{\;\;a}^{\mu },\;\mathrm{antigh}%
\left( b_{a}\right) =
2,\;\mathrm{antigh}\left( c_{\;\;a}^{\mu }\right) =1.
\label{4.14d}
\end{equation}
The currents associated with a trivial
global symmetry are trivial \cite{20}%
, and this case will be not considered here.
Inserting (\ref{4.14a}) in (\ref%
{4.14}), we find that
\begin{equation}
a_{0}=\frac{1}{2}f_{\;\;ab}^{c}
F_{c}^{\mu \nu }A_{\mu }^{a}A_{\nu
}^{b}+j_{\;\;a}^{\mu }A_{\mu }^{a}.
\label{4.18}
\end{equation}
So far, we have constructed the first-order
deformation of the solution to
the master equation in the form
\begin{eqnarray}
& &S_{1}=\int d^{D}x\left( \frac{1}{2}
f_{\;\;ab}^{c}F_{c}^{\mu \nu }A_{\mu
}^{a}A_{\nu }^{b}+j_{\;\;a}^{\mu }A_{\mu }^{a}+\right. .
\nonumber \\
&&\left. f_{\;\;ab}^{c}A_{c}^{*\mu }A_{\mu }^{b}\eta
^{a}+y_{i}^{*}T_{\;\;a}^{i}\eta ^{a}-
\frac{1}{2}f_{\;\;ab}^{c}\eta
_{c}^{*}\eta ^{a}\eta ^{b}\right) .  \label{4.19}
\end{eqnarray}
The appearance of the second term in
$S_{1}$ emphasizes that the first-order
coupling among vector fields and matter
fields is realized at the level of
the deformed Lagrangian through an
interaction term of the type $%
gj_{\;\;a}^{\mu }A_{\mu }^{a}$, that
connects them precisely through the
conserved currents associated with the
global symmetries of the matter
theory.

\subsection{Higher-order deformations}

In the following we approach the
higher-order consistency of the
deformation. The eq. that governs the
second-order deformation, (\ref{2.6}),
demands that the antibracket
\begin{eqnarray}
&&\left( S_{1},S_{1}\right) =
s\left( \int d^{D}x\frac{1}{2}f_{\;%
\;ab}^{c}f_{cde}A_{\mu }^{a}
A_{\nu }^{b}A^{d\mu }A^{e\nu }\right) +
\nonumber \\
&&\int d^{D}x\left( 2\left(
f_{\;\;ab}^{c}j_{\;\;c}^{\mu }+\frac{\delta
^{R}j_{\;\;b}^{\mu }}{\delta y^{i}}
T_{\;\;a}^{i}\right) A_{\mu }^{b}\eta
^{a}-\right. \nonumber \\
& &f_{\;\;b\left[ a\right. }^{c}
f_{\;\;\left. de\right] }^{b}\left( F_{c}^{\mu
\nu }A_{\mu }^{a}A_{\nu }^{e}
\eta ^{d}+A_{c}^{*\mu }A_{\mu }^{e}\eta
^{a}\eta ^{d}+\frac{1}{3}
\eta _{c}^{*}\eta ^{a}\eta ^{d}\eta ^{e}\right) +
\nonumber \\
&&\left. y_{i}^{*}\left( \frac{\delta ^{R}
T_{\;\;a}^{i}}{\delta y^{j}}%
T_{\;\;b}^{j}-\frac{\delta ^{R}
T_{\;\;b}^{i}}{\delta y^{j}}%
T_{\;\;a}^{j}-f_{\;\;ab}^{c}T_{\;\;c}^{i}\right) 
\eta ^{a}\eta ^{b}\right) ,
\label{4.21}
\end{eqnarray}
is a $s$-coboundary, or, equivalently,
that its integrand is $s$-exact
modulo $d$. As the last two terms in
(\ref{4.21}) cannot be $s$-exact modulo
$d$, it follows that they should be made
to vanish. This holds if and only
if
\begin{equation}
f_{\;\;b\left[ a\right. }^{c}
f_{\;\;\left. de\right] }^{b}=0,  \label{4.21a}
\end{equation}
\begin{equation}
\frac{\delta ^{R}T_{\;\;a}^{i}}{\delta y^{j}}
T_{\;\;b}^{j}-\frac{\delta
^{R}T_{\;\;b}^{i}}{\delta y^{j}}T_{\;\;a}^{j}=
f_{\;\;ab}^{c}T_{\;\;c}^{i}.
\label{4.21b}
\end{equation}
On the one hand, the relation (\ref{4.21a})
shows that the antisymmetric
constants $f_{\;\;ab}^{c}$ satisfy Jacobi's
identity, and hence they stand
for the structure constants of a Lie algebra.
On the other hand, the
solution to the eq. (\ref{4.21b}) can
be expressed in the form
\begin{equation}
T_{\;\;a}^{i}=T_{\;\;ak}^{i}y^{k},  \label{4.21c}
\end{equation}
where $T_{\;\;ak}^{i}$ denotes a basis
of generators of the linear
representation of the gauge algebra
under which the matter fields $y^{i}$
transform, i.e.
\begin{equation}
T_{\;\;aj}^{i}T_{\;\;bk}^{j}-T_{\;\;bj}^{i}
T_{\;\;ak}^{j}=f_{\;\;ab}^{c}T_{%
\;\;ck}^{i}.  \label{4.21d}
\end{equation}
From the last component of (\ref{4.19}),
as well as from the Jacobi identity
(\ref{4.21a}), it follows that the deformed
gauge transformations of the
vector fields form a Lie algebra with the
structure constants $f_{\;\;ab}^{c}
$. In the meantime, the relations
(\ref{4.21b}-\ref{4.21d}) express that the
deformed gauge transformations of the
matter fields also generate a Lie
algebra with the same structure constants.
In conclusion, the entire
deformed gauge algebra is Lie.
Substituting (\ref{4.21a}-\ref{4.21b}) in (%
\ref{4.21}), we derive
\begin{eqnarray}
&&\left( S_{1},S_{1}\right) =
s\left( \int d^{D}x\frac{1}{2}f_{\;%
\;ab}^{c}f_{cde}A_{\mu }^{a}A_{\nu }^{b}
A^{d\mu }A^{e\nu }\right) +
\nonumber \\
&&\int d^{D}x2\left( f_{\;\;ab}^{c}
j_{\;\;c}^{\mu }+\frac{\delta
^{R}j_{\;\;b}^{\mu }}{\delta y^{i}}
T_{\;\;a}^{i}\right) A_{\mu }^{b}\eta
^{a}.  \label{4.21e}
\end{eqnarray}

Looking at (\ref{4.21e}), we observe
that two main cases arise. First, it
might happen that the currents transform
under the gauge version of the
rigid symmetries of the matter fields
according to the adjoint
representation of the Lie gauge algebra,
hence they satisfy the relations
\begin{equation}
f_{\;\;ab}^{c}j_{\;\;c}^{\mu }+
\frac{\delta ^{R}j_{\;\;b}^{\mu }}{\delta
y^{i}}T_{\;\;a}^{i}=0.  \label{4.22}
\end{equation}
This further leads to $\left( S_{1},S_{1}\right) =
s\left( \int d^{D}x\frac{1%
}{2}f_{\;\;ab}^{c}f_{cde}A_{\mu }^{a}
A_{\nu }^{b}A^{d\mu }A^{e\nu }\right) $%
, such that
\begin{equation}
S_{2}=-\frac{1}{4}\int d^{D}x
f_{\;\;ab}^{c}f_{cde}A_{\mu }^{a}A_{\nu
}^{b}A^{d\mu }A^{e\nu }.  \label{4.22a}
\end{equation}
As $\left( S_{1},S_{2}\right) =0$, we
can safely take all the higher-order
deformations equal to zero, $S_{3}=S_{4}=
\cdots =0$, in which case the
deformed solution to the master equation
consistent to all orders in the
coupling constant is given by
\begin{eqnarray}
&&S=\int d^{D}x\left( -\frac{1}{4}
\tilde{F}_{\mu \nu }^{a}\tilde{F}_{a}^{\mu
\nu }+{\mathcal{L}}_{0}\left( y^{i},
\partial _{\mu }y^{i},\ldots ,\partial
_{\mu _{1}}\cdots \partial _{\mu _{k}}
y^{i}\right) +gj_{\;\;a}^{\mu }A_{\mu
}^{a}+\right.  \nonumber \\
&&\left. A_{a}^{*\mu }\left(
D_{\mu }\right) _{\;\;b}^{a}\eta
^{b}+gy_{i}^{*}T_{\;\;ak}^{i}y^{k}
\eta ^{a}-\frac{1}{2}gf_{\;\;ab}^{c}\eta
_{c}^{*}\eta ^{a}\eta ^{b}\right) ,  \label{4.23}
\end{eqnarray}
where
\begin{equation}
\tilde{F}_{\mu \nu }^{a}=\partial _{\mu }
A_{\nu }^{a}-\partial _{\nu }A_{\mu
}^{a}-gf_{\;\;bc}^{a}A_{\mu }^{b}A_{\nu }^{c},
\label{4.23a}
\end{equation}
\begin{equation}
\left( D_{\mu }\right) _{\;\;b}^{a}=
\delta _{\;\;b}^{a}\partial _{\mu
}+gf_{\;\;bc}^{a}A_{\mu }^{c},  \label{4.23b}
\end{equation}
represents the field strength of
non-abelian gauge fields, respectively, the
covariant derivative in the adjoint
representation. From the
antifield-independent piece in (\ref{4.23})
we read that the overall
Lagrangian action of the interacting gauge
theory has the expression
\begin{equation}
\bar{S}=\int d^{D}x\left( -\frac{1}{4}
\tilde{F}_{\mu \nu }^{a}\tilde{F}%
_{a}^{\mu \nu }+{\mathcal{L}}_{0}
\left( y^{i},\partial _{\mu }y^{i},\ldots
,\partial _{\mu _{1}}\cdots
\partial _{\mu _{k}}y^{i}\right)
+gj_{\;\;a}^{\mu }A_{\mu }^{a}\right) ,  \label{4.24}
\end{equation}
while from the components linear in the
antifields we conclude that it is
invariant under the gauge transformations
\begin{equation}
\bar{\delta}_{\epsilon }A_{\mu }^{a}=
\left( D_{\mu }\right)
_{\;\;b}^{a}\epsilon ^{b},\;\bar{\delta}_{\epsilon
}y^{i}=gT_{\;\;ak}^{i}y^{k}\epsilon ^{a}.  \label{4.20}
\end{equation}
Thus, if the currents present in the purely
matter theory transform under
the gauge version of the rigid symmetries of
the matter fields according to
the adjoint representation of the Lie gauge
algebra, then the only
interacting term that couples the vector
fields to the matter fields is of
the type $j_{\;\;a}^{\mu }A_{\mu }^{a}$.

In the opposite case, where the currents do
not satisfy (\ref{4.22}),
\begin{equation}
f_{\;\;ab}^{c}j_{\;\;c}^{\mu }+
\frac{\delta ^{R}j_{\;\;b}^{\mu }}{\delta
y^{i}}T_{\;\;a}^{i}\neq 0,  \label{4.25}
\end{equation}
it follows that the second term from the
right hand-side of (\ref{4.21e}) is
non-vanishing, hence the second-order
deformation involving vector fields
coupled to matter fields will also be
non-trivial. Moreover, it is possible
to obtain other non-trivial higher-order
deformations when solving the
appropriate eqs., whose expressions
depend on the structure of the matter
theory, and hence cannot be output in
the general case discussed so far.
Anyway, the complete deformed solution
to the master equation starts like
\begin{equation}
S^{^{\prime }}=S+g^{2}S_{2}+\cdots ,  \label{4.26}
\end{equation}
such that the Lagrangian action of the
interacting theory is of the type
\begin{equation}
\bar{S}^{^{\prime }}=\bar{S}+
{\mathcal{O}}\left( g^{2}\right) .  \label{4.27}
\end{equation}
In this situation, the gauge
transformations of the fields involved with the
interacting theory are also given by
(\ref{4.20}).

The deformation treatment developed so
far can be synthesized in three
general results as follows. First, the
interaction terms involving only the
vector fields generate the Lagrangian
action of Yang-Mills theory, and the
first-order couplings between the vector
fields and matter fields is of the
type $j_{\;\;a}^{\mu }A_{\mu }^{a}$.
Second, the gauge transformations of
the vector fields are modified with
respect to the initial ones, while those
of the matter fields can be obtained by
simply gauging the original rigid
ones. Third, the overall deformed gauge
algebra is a Lie algebra. Finally, a
word of caution. Once the deformations
related to a given matter theory are
computed, special attention should be
paid to the elimination of
non-locality, as well as of triviality
of the resulting deformations. This
completes our general procedure.

\section{Applications}

Next, we consider two examples of
matter theories --- real scalar fields and
Dirac fields --- and compute their
consistent interactions with a set of
vector fields in the light of the
analysis performed in the previous section.

\subsection{Vector fields coupled to scalar fields}

First, we analyse the consistent interactions
that can be introduced among a
set of real scalar fields and a collection of
vector fields. In this case
the ``free'' Lagrangian action
(\ref{4.1}) is given by
\begin{equation}
\tilde{S}_{0}^{L}\left[ A_{\mu }^{a},\varphi ^{A}\right] =
\int d^{4}x\left( -%
\frac{1}{4}F_{\mu \nu }^{a}F_{a}^{\mu \nu }+
\frac{1}{2}K_{AB}\left( \partial
_{\mu }\varphi ^{A}\right) \left( 
\partial ^{\mu }\varphi ^{B}\right)
-V\left( \varphi ^{A}\right) \right) ,  \label{5.1}
\end{equation}
where $K_{AB}$ is an invertible symmetric
constant matrix. The ``free''
Koszul-Tate differential and the exterior
derivative along the gauge orbits
act on the generators associated with the
matter sector $y^{i}=\left(
\varphi ^{A}\right) $,
$y_{i}^{*}=\left( \varphi _{A}^{*}\right) $ like
\begin{equation}
\delta \varphi ^{A}=0,\;\gamma \varphi ^{A}=0,
\label{5.2}
\end{equation}
\begin{equation}
\delta \varphi _{A}^{*}=K_{AB}
\partial _{\mu }\partial ^{\mu }\varphi ^{B}+%
\frac{\partial V}{\partial
\varphi ^{A}},\;\gamma \varphi _{A}^{*}=0,
\label{5.3}
\end{equation}
while on the purely gauge ones
$A_{\mu }^{a}$, $\eta ^{a}$, $A_{a}^{*\mu }$
and $\eta _{a}^{*}$ like in the first
and third relations from each of the
eqs. (\ref{4.5}--\ref{4.8}).
Multiplying (\ref{5.3}) by $-T_{\;\;aC}^{A}%
\varphi ^{C}$, we arrive at the eq.
\begin{eqnarray}
&&\delta \left( -\varphi _{A}^{*}
T_{\;\;aC}^{A}\varphi ^{C}\right) =\partial
_{\mu }\left( -K_{AB}T_{\;\;aC}^{A}\left(
\partial ^{\mu }\varphi
^{B}\right) \varphi ^{C}\right) +
\nonumber \\
& &K_{AB}T_{\;\;aC}^{A}\left(
\partial ^{\mu }\varphi ^{B}\right) \left(
\partial _{\mu }\varphi ^{C}\right) -
\frac{\partial V}{\partial \varphi ^{A}}%
T_{\;\;aC}^{A}\varphi ^{C}.  \label{5.6}
\end{eqnarray}
In order to have a non-trivial local
homology for $\delta $ at antighost
number one, the right hand-side of
(\ref{5.6}) should reduce to a total
derivative. This takes place if and only if
\begin{equation}
\frac{\partial V}{\partial \varphi ^{A}}
T_{\;\;aC}^{A}\varphi ^{C}=0,
\label{5.6a}
\end{equation}
\begin{equation}
\tilde{T}_{\;\;aBC}=-\tilde{T}_{\;\;aCB},
\label{5.6b}
\end{equation}
with $\tilde{T}_{\;\;aBC}\equiv
K_{AB}T_{\;\;aC}^{A}$. The equations (\ref%
{5.6a}-\ref{5.6b}) ensure the invariance
of the action of the scalar theory
under the rigid transformations
$\Delta \varphi ^{A}=T_{\;\;aC}^{A}\varphi
^{C}\xi ^{a}$, such that we obtain
the conservation of the currents
\begin{equation}
j_{\;\;a}^{\mu }=-K_{AB}T_{\;\;aC}^{A}
\left( \partial ^{\mu }\varphi
^{B}\right) \varphi ^{C}.  \label{5.7}
\end{equation}
Then, with the help of (\ref{4.19})
and (\ref{5.7}), the first-order
deformation of the solution to the
master equation reads
\begin{eqnarray}
& &S_{1}=\int d^{4}x\left( \frac{1}{2}
f_{\;\;ab}^{c}F_{c}^{\mu \nu }A_{\mu
}^{a}A_{\nu }^{b}-K_{AB}T_{\;\;aC}^{A}
\left( \partial ^{\mu }\varphi
^{B}\right) \varphi ^{C}A_{\mu }^{a}+\right. .
\nonumber \\
&&\left. f_{\;\;ab}^{c}A_{c}^{*\mu }
A_{\mu }^{b}\eta ^{a}+\varphi
_{A}^{*}T_{\;\;aC}^{A}\varphi ^{C}\eta ^{a}-
\frac{1}{2}f_{\;\;ab}^{c}\eta
_{c}^{*}\eta ^{a}\eta ^{b}\right) .  \label{5.12}
\end{eqnarray}
Now, we investigate the second-order
deformation. In view of this, we
observe that the second term in the
right hand-side of (\ref{4.21e}) adapted
to the model under study is given by
\begin{equation}
2\left( f_{\;\;ab}^{c}j_{\;\;c}^{\mu }+
\frac{\delta ^{R}j_{\;\;b}^{\mu }}{%
\delta \varphi ^{A}}T_{\;\;aC}^{A}
\varphi ^{C}\right) A_{\mu }^{b}\eta
^{a}=s\left( K_{AB}T_{\;\;aD}^{A}
T_{\;\;bC}^{D}\varphi ^{B}\varphi
^{C}A_{\mu }^{a}A^{b\mu }\right) ,  \label{5.12a}
\end{equation}
such that
\begin{eqnarray}
& &S_{2}=\int d^{D}x\left( -
\frac{1}{4}f_{\;\;ab}^{c}f_{cde}A_{\mu }^{a}A_{\nu
}^{b}A^{d\mu }A^{e\nu }+\right.
\nonumber \\
&&\left. \frac{1}{2}K_{AD}
T_{\;\;aB}^{A}T_{\;\;bC}^{D}\varphi ^{B}\varphi
^{C}A_{\mu }^{a}A^{b\mu }\right) .  \label{5.15}
\end{eqnarray}
This case corresponds to the second
situation described in the final part of
Sec. 3. If we examine the third-order
deformation (see eq. (\ref{2.7})), we
notice that $\left( S_{1},S_{2}\right) =0$,
hence we can safely take $S_{3}=0
$. The higher-order eqs. are then
satisfied with the choice $%
S_{4}=S_{5}=\cdots =0$. Putting together
the above results, we infer that
\begin{eqnarray}
&&S^{\prime }=\int d^{4}x\left( -
\frac{1}{4}\tilde{F}_{\mu \nu }^{a}\tilde{F}%
_{a}^{\mu \nu }+\frac{1}{2}K_{AB}
\left( D_{\mu \;\;C}^{A}\varphi ^{C}\right)
\left( D_{\;\;\;D}^{\mu B}\varphi ^{D}\right) -
V\left( \varphi ^{A}\right)
+\right.   \nonumber \\
&&\left. g\varphi _{A}^{*}
T_{\;\;aC}^{A}\varphi ^{C}\eta ^{a}+A_{a}^{*\mu
}\left( D_{\mu }\right) _{\;\;b}^{a}
\eta ^{b}-\frac{1}{2}gf_{\;\;ab}^{c}\eta
_{c}^{*}\eta ^{a}\eta ^{b}\right) ,  \label{5.16}
\end{eqnarray}
represents the full consistent solution
to the master equation of our
deformed problem, where the covariant
derivative that acts on the matter
fields is defined through
\begin{equation}
D_{\mu \;\;C}^{A}=\delta _{\;\;C}^{A}\partial _{\mu }+
gT_{\;\;aC}^{A}A_{\mu
}^{a}.  \label{5.17}
\end{equation}
The antifield-independent piece in (\ref{5.16})
\begin{equation}
\bar{S}^{\prime }=\int d^{4}x\left( -
\frac{1}{4}\tilde{F}_{\mu \nu }^{a}%
\tilde{F}_{a}^{\mu \nu }+\frac{1}{2}
K_{AB}\left( D_{\mu \;\;C}^{A}\varphi
^{C}\right) \left( D_{\;\;\;D}^{\mu B}
\varphi ^{D}\right) -V\left( \varphi
^{A}\right) \right) ,  \label{5.18}
\end{equation}
describes nothing but the Lagrangian
interaction between a set of real
scalar fields and a collection of vector
fields, while the terms linear in
the antifields of the matter fields give
the gauge transformations
\begin{equation}
\bar{\delta}_{\epsilon }\varphi ^{A}=
gT_{\;\;aC}^{A}\varphi ^{C}\epsilon
^{a},  \label{5.19}
\end{equation}
those associated with the vector fields
being as in the former relations
from (\ref{4.20}). If in
(\ref{5.18}--\ref{5.19}) we make the transformation
$T_{\;\;aC}^{A}\rightarrow iT_{\;\;aC}^{A}$,
we arrive at the non-abelian
analogue of scalar electrodynamics.

\subsection{Vector fields coupled to Dirac fields}

Finally, we examine the consistent couplings
between a set of vector fields
and a collection of massive Dirac fields.
In view of this, we start from the
Lagrangian action
\begin{equation}
\tilde{S}_{0}^{L}\left[ A_{\mu }^{a},
\psi _{\;A}^{\alpha },\bar{\psi}%
_{\alpha }^{\;A}\right] =\int d^{4}x\left( -
\frac{1}{4}F_{\mu \nu
}^{a}F_{a}^{\mu \nu }+
\bar{\psi}_{\alpha }^{\;A}\left( i\left( \gamma ^{\mu
}\right) _{\;\;\beta }^{\alpha }\partial _{\mu }-
m\delta _{\;\;\beta
}^{\alpha }\right) \psi _{\;A}^{\beta }\right) ,
\label{5.20}
\end{equation}
where $\psi _{\;A}^{\alpha }$ and
$\bar{\psi}_{\alpha }^{\;A}$ denote the
fermionic spinor components of the
Dirac fields $\psi _{\;A}$ and $\bar{\psi}%
^{\;A}$. The actions of $\delta $ and
$\gamma $ on the matter generators
from the free BRST complex are expressed by
\begin{equation}
\delta \psi _{\;A}^{\alpha }=
0,\;\delta \bar{\psi}_{\alpha }^{\;A}=0,
\label{5.21}
\end{equation}
\begin{equation}
\delta \psi _{\alpha }^{*A}=
-\left( i\left( \gamma ^{\mu }\right)
_{\;\;\alpha }^{\beta }\partial _{\mu }+
m\delta _{\;\;\alpha }^{\beta
}\right) \bar{\psi}_{\beta }^{\;A},
\label{5.22}
\end{equation}
\begin{equation}
\delta \bar{\psi}_{A}^{*\alpha }=
-\left( i\left( \gamma ^{\mu }\right)
_{\;\;\beta }^{\alpha }\partial _{\mu }-
m\delta _{\;\;\beta }^{\alpha
}\right) \psi _{\;A}^{\beta },  \label{5.23}
\end{equation}
\begin{equation}
\gamma \psi _{\;A}^{\alpha }=0,\;\gamma 
\bar{\psi}_{\alpha
}^{\;A}=0,\;\gamma \psi _{\alpha }^{*A}=
0,\;\gamma \bar{\psi}_{A}^{*\alpha
}=0,  \label{5.24}
\end{equation}
where the antighost number one antifields
$\psi _{\alpha }^{*A}$ and $\bar{%
\psi}_{A}^{*\alpha }$ are bosonic.
Multiplying (\ref{5.22}) and (\ref{5.23})
from the right by $T_{\;\;aA}^{C}\psi _{\;C}^{\alpha }$,
respectively, $%
-T_{\;\;aC}^{A}\bar{\psi}_{\alpha }^{\;C}$,
and subtracting the resulting
relations, we deduce
\begin{equation}
\delta \left( \psi _{\alpha }^{*A}T_{\;\;aA}^{C}
\psi _{\;C}^{\alpha }-\bar{%
\psi}_{A}^{*\alpha }T_{\;\;aC}^{A}
\bar{\psi}_{\alpha }^{\;C}\right)
=\partial _{\mu }\left( 
i\bar{\psi}_{\alpha }^{\;A}\left( \gamma ^{\mu
}\right) _{\;\;\beta }^{\alpha }
T_{\;\;aA}^{C}\psi _{\;C}^{\beta }\right) ,
\label{5.25}
\end{equation}
which underlines the conservation of the currents
\begin{equation}
j_{\;\;a}^{\mu }=
i\bar{\psi}_{\alpha }^{\;A}\left( \gamma ^{\mu }\right)
_{\;\;\beta }^{\alpha }T_{\;\;aA}^{C}
\psi _{\;C}^{\beta },  \label{5.26}
\end{equation}
resulting from the global invariances
$\Delta \psi _{\;A}^{\alpha
}=-T_{\;\;aA}^{C}\psi _{\;C}^{\alpha }
\xi ^{a}$ and $\Delta \bar{\psi}%
_{\alpha }^{\;A}=T_{\;\;aC}^{A}
\bar{\psi}_{\alpha }^{\;C}\xi ^{a}$ of the
massive Dirac theory. Acting like before,
the first-order deformed solution
to the master equation is given by
\begin{eqnarray}
&&S_{1}=\int d^{4}x\left( \frac{1}{2}
f_{\;\;ab}^{c}F_{c}^{\mu \nu }A_{\mu
}^{a}A_{\nu }^{b}+
i\bar{\psi}_{\alpha }^{\;A}\left( \gamma ^{\mu }\right)
_{\;\;\beta }^{\alpha }T_{\;\;aA}^{C}
\psi _{\;C}^{\beta }A_{\mu
}^{a}+\right. 
\nonumber \\
&&\left. f_{\;\;ab}^{c}A_{c}^{*\mu }
A_{\mu }^{b}\eta ^{a}-\left( \psi _{\alpha
}^{*A}T_{\;\;aA}^{C}\psi _{\;C}^{\alpha }-
\bar{\psi}_{A}^{*\alpha
}T_{\;\;aC}^{A}\bar{\psi}_{\alpha }^{\;C}\right)
\eta ^{a}-\frac{1}{2}%
f_{\;\;ab}^{c}\eta _{c}^{*}\eta ^{a}
\eta ^{b}\right) .  \label{5.31}
\end{eqnarray}
With this solution at hand, we observe
that the second term in the right
hand-side of (\ref{4.21e}) adapted to
the model under study vanishes
\begin{equation}
f_{\;\;ab}^{c}j_{\;\;c}^{\mu }+
\frac{\delta ^{R}j_{\;\;b}^{\mu }}{\delta
\bar{\psi}_{\alpha }^{\;A}}
T_{\;\;aB}^{A}\bar{\psi}_{\alpha }^{\;B}-\frac{%
\delta ^{R}j_{\;\;b}^{\mu }}{\delta
\psi _{\;A}^{\alpha }}T_{\;\;aA}^{B}\psi
_{\;B}^{\alpha }=0,  \label{5.31a}
\end{equation}
so $S_{2}$ is given by (\ref{4.22a}).
In this light, we conclude that the
present model satisfies the conditions
described in the first situation
discussed in the final part of Sec. 3.
Consequently, we can set $S_{3}=0$,
and, moreover, $S_{4}=S_{5}=\cdots =0$.
The complete deformed solution to
the master equation that is consistent to
all orders in the deformation
parameter is
\begin{eqnarray}
&&S=\int d^{4}x\left( -
\frac{1}{4}\tilde{F}_{\mu \nu }^{a}\tilde{F}_{a}^{\mu
\nu }+\bar{\psi}_{\alpha }^{\;A}\left( 
i\left( \gamma ^{\mu }\right)
_{\;\;\beta }^{\alpha }D_{\mu \;\;A}^{C}-
m\delta _{\;\;\beta }^{\alpha
}\delta _{\;\;A}^{C}\right) \psi _{\;C}^{\beta }+
\right.   \nonumber \\
&&A_{a}^{*\mu }\left(
D_{\mu }\right) _{\;\;b}^{a}\eta ^{b}-g\left(
\psi _{\alpha }^{*A}T_{\;\;aA}^{C}
\psi _{\;C}^{\alpha }-\bar{\psi}%
_{A}^{*\alpha }T_{\;\;aC}^{A}
\bar{\psi}_{\alpha }^{\;C}\right) \eta ^{a}-%
\nonumber \\
&&\left. \frac{1}{2}gf_{\;\;ab}^{c}
\eta _{c}^{*}\eta ^{a}\eta ^{b}\right) .
\label{5.33}
\end{eqnarray}
Its antifield-independent piece
\begin{equation}
\bar{S}=\int d^{4}x\left( -
\frac{1}{4}\tilde{F}_{\mu \nu }^{a}\tilde{F}%
_{a}^{\mu \nu }+
\bar{\psi}_{\alpha }^{\;A}\left( i\left( \gamma ^{\mu
}\right) _{\;\;\beta }^{\alpha }D_{\mu \;\;A}^{C}-
m\delta _{\;\;\beta
}^{\alpha }\delta _{\;\;A}^{C}\right)
\psi _{\;C}^{\beta }\right) ,
\label{5.34}
\end{equation}
reveals the Lagrangian interaction
between a set of vector fields and a
collection of massive Dirac fields,
while from the terms linear in the
antifields of the matter fields we
notice the gauge transformation of the
Dirac fields are
\begin{equation}
\delta _{\epsilon }\psi _{\;A}^{\alpha }=
-gT_{\;\;aA}^{C}\psi _{\;C}^{\alpha
}\epsilon ^{a},\;\delta _{\epsilon }
\bar{\psi}_{\alpha
}^{\;A}=gT_{\;\;aC}^{A}
\bar{\psi}_{\alpha }^{\;C}\epsilon ^{a},  \label{5.35}
\end{equation}
those associated with the vector
fields being as in the former relations
from (\ref{4.20}). Similar to the
scalar case, if in (\ref{5.34}--\ref{5.35}%
) we make the transformation
$T_{\;\;aC}^{A}\rightarrow iT_{\;\;aC}^{A}$ and
consider an $SU\left( 3\right) $
gauge algebra, we arrive at quantum
chromodynamics.

\section{Conclusion}

In conclusion, in this paper we
have investigated the consistent
interactions that can be introduced
between a set of vector fields and a
system of matter fields by using some
cohomological techniques. In this
context, we have shown that the
first-order interaction contains two terms.
The first one describes an interaction
among the vector fields (the cubic
vertex of Yang-Mills theory). The
second term appears only if the matter
theory displays as many global
symmetries as there are vector fields, and is
of the form $j_{\;\;a}^{\mu }A_{\mu }^{a}$,
where $j_{\;\;a}^{\mu }$ are the
conserved currents of the matter theory
corresponding to the above mentioned
global symmetries. The consistency of
the first-order deformation requires
that the deformed gauge algebra is Lie,
and, in the meantime, emphasizes
that the second-order deformation
contains the quartic vertex of pure
Yang-Mills theory. If the conserved
currents $j_{\;\;a}^{\mu }$ transform
under the gauge version of the rigid
symmetries of the matter fields
according to the adjoint representation
of the Lie gauge algebra, then all
the other deformations involving the matter
fields, of order two and higher,
vanish. In the opposite case, at least the
second-order deformation implying
matter fields is non-vanishing, but in
principle there might be other
non-trivial terms. The general procedure
has been applied to the study of
the interactions between a set of vector
fields and a collection of real
scalar fields, respectively, a set
of Dirac fields.

\section*{Acknowledgment}

This work has been supported by a Romanian
National Council for Academic
Scientific Research (CNCSIS) grant.

\end{document}